\documentclass[sigconf]{acmart}
\acmConference[EASE 2025]{The 29th International Conference on Evaluation and Assessment in Software Engineering}{17–20 June, 2025}{Istanbul, Türkiye}
\usepackage{natbib}
\usepackage{algorithmic}
\usepackage{graphicx}
\usepackage{textcomp}
\usepackage{xcolor}
\usepackage{booktabs}
\usepackage{soul}
\usepackage{xcolor}
\sethlcolor{yellow} % or any other color like lightgray, cyan, etc.

\usepackage{multirow}
\def\BibTeX{{\rm B\kern-.05em{\sc i\kern-.025em b}\kern-.08em
    T\kern-.1667em\lower.7ex\hbox{E}\kern-.125emX}}
\usepackage{url}

\usepackage{framed}
\usepackage{changepage}

\newcommand{\interviewquote}[2]{
 \def\FrameCommand{%
    \hspace{0pt}%
    {\color{brown} \vrule width 2pt}% <-- Change color here.
    \colorbox{white}
  }%
  \MakeFramed{\advance\hsize-\width\FrameRestore}%
  \noindent% disable indenting first paragraph
  \begin{adjustwidth}{}{1pt}
  {\small``\textit{#1}'' - {#2}}\end{adjustwidth}\endMakeFramed%
}

\begin{document}

\title{Evaluating the Impact of a Yoga-Based Intervention on Software Engineers' Well-Being}

\author{Cristina Martinez Montes}
\orcid{0000-0003-1150-6931}
\affiliation{%
  \institution{Chalmers University of Technology and University of Gothenburg}
  \city{Gothenburg}
  \country{Sweden}
}
\email{montes@chalmers.se}

\author{Birgit Penzenstadler}
\orcid{0000-0002-5771-0455}
\affiliation{%
  \institution{Chalmers University of Technology and University of Gothenburg}
  \city{Gothenburg}
  \country{Sweden}
}

\affiliation{
   \institution{Lappeenranta University of Technology}
   \city{Lappeenranta}
    \country{Finland}
}
\email{birgitp@chalmers.se}

\begin{abstract}

%Background
Software engineering tasks are high-stress and cognitively demanding. Additionally, there is a latent risk of software engineers presenting burnout, depression and anxiety. Established interventions in other fields centred around attention awareness have shown positive results in mental well-being.

We aim to test how effective a yoga intervention is in improving general well-being in the workplace. For that, we designed, implemented and evaluated an eight-week yoga programme in a software development company. We used a mixed-methods data collection, using a survey of six psychometric scales, pre and post-intervention, and a weekly well-being scale during the programme. For method triangulation, we conducted a focus group with the organisers to obtain qualitative data.
The quantitative results did not show any statistically significant improvement after the intervention. Meanwhile, the qualitative results illustrated that participants felt better and liked the intervention. 

We conclude that yoga has a positive impact, which, however, can easily get overlaid by contextual factors, especially with only a once-per-week intervention.

\end{abstract}


\maketitle
\section{Introduction} \label{sec:intro}

%Background Information
Stress is an increasing concern in modern society, with work-related stress being particularly prevalent in high-demand environments. This is especially true in fields like software engineering, where mental workload, strict deadlines, and extended periods of sedentary behaviour contribute to burnout \cite{maudgalya2006workplace}, and reduced well-being \cite{wong2023mental}. 

Work-related stress is detrimental to workers' psychological health and costly to societies. A broad analysis by the American Institute of Stress \cite{kern2022workstress}, factoring in absenteeism, turnover, reduced productivity, and higher medical and legal expenses, raised the estimate to \$300 billion annually. Regarding Europe, Shaholli et al. \cite{shaholli2023work} reviewed international studies and organisational reports to estimate the economic impact of occupational stress. Their findings reveal estimates ranging from €54 million to €280 billion, depending on the country.

Mindfulness practices have proven beneficial in demanding, high-stress work settings that require intense focus. Leading tech companies, including Intel, Goldman Sachs, Google, and SAP, have widely embraced it to promote employee well-being \cite{good2016contemplating, schultz2015mindfulness, everson2015sap}.

%Problem Statement / missing knowledge

Mindfulness-based programmes have been implemented in different contexts, generally getting positive results. Penzenstadler et al.~\cite{penzenstadler2022take} did an online intervention using breathwork to improve general well-being and reduce stress in participants. Their results were positive, with an increase in attention and positive thinking. Further, Montes et al. \cite{montes2024qualifying} elaborate from a qualitative perspective on a similar intervention, sharing participants' positive perceptions of their course. Few interventions are done in the context of software engineering workplaces; for example, Bernardez~\cite {bernardez2023empirical} studied the effect of mindfulness practice, meditation, on a sample of 56 helpdesk employees working for a consulting and information technology company. Their participants significantly improved attention awareness. Heijer et al.~\cite {den2017don} studied the impact of mindfulness on agile software teams in over two months of stand-up meetings with 61 participants from eight companies. The findings showed improved perceived effectiveness, decision-making, and listening. 

%research significance
This study is among the first ones carried out in a workplace setting with a software engineer population and using standardised scales to measure the effects of yoga as a mindfulness practice.

%Specific Focus and RQs

In this study, we aim to answer the following research question: 

\textbf{How does a workplace yoga intervention impact the general well-being of software engineers?} 

We approach this through a quasi-experiment mixed-methods design, using psychometric instruments to measure pre- and post-intervention well-being. We use six psychometric scales complemented by qualitative data from focus groups to provide deeper insights into the participants' experiences.

%paper overview (if enough space)

The paper is organised as follows: Section II reviews the related work on the benefits of yoga, specifically Hatha yoga, software engineers' well-being and existing interventions. Section III outlines the study's methodology, including participant recruitment, intervention design, and data collection and analysis procedures. Section IV presents the findings, and Section V discusses the results, limitations, and implications for practice. Finally, Section VI concludes the study and offers directions for future research.
\section{Related Work} \label{sec:related}

Yoga is an ancient Indian practice designed to ``still the fluctuations of the mind'' and facilitate meditative absorption, a psychological state marked by feelings of self-transcendence and unceasing happiness~\cite{bryant2015yoga}. A regular yoga practice can improve strength, flexibility, and balance; reduce stress; and provide many therapeutic benefits~\cite{markil2010hatha}.
The most common style of yoga practiced in Western countries is \textbf{Hatha yoga}, which includes synchronized movements through postures with breath, meditation, breathing exercises, and supine rest to conclude~\cite{luu2016hatha}.
Hatha Yoga is classified as a mind-body exercise (along with Tai Chi, Qi Gong, Pilates, and others) and a type of complementary and alternative medicine that has become a popular and effective form of exercise because of the numerous health and fitness benefits associated with a regular practice.\footnote{\url{https://www.nccih.nih.gov/health/yoga-effectiveness-and-safety}}

\subsection{The Effectiveness of Yoga in General}

A number of meta studies has collected evidence on the positive effects of yoga practice:

Ross et al.~\cite{ross2010health} used the key word ``yoga,'' on PubMed and yielded 81 studies that met inclusion criteria. These studies subsequently were classified as uncontrolled (n=30), wait list controlled (n=16), or comparison (n=35). The most common comparison intervention (n=10) involved exercise. In the studies reviewed, yoga interventions appeared to be equal or superior to exercise in nearly every outcome measured except those involving physical fitness. The studies comparing the effects of yoga and exercise seem to indicate that, in both healthy and diseased populations, yoga may be as effective as or better than exercise at improving a variety of health-related outcome measures~\cite{ross2010health}. This empirical evidence is important to show the feasibility and likelihood of success of using yoga as mode of intervention in the study at hand.

Cramer et al.~\cite{cramer2013yoga} searched Medline/PubMed, Scopus, the Cochrane Library, PsycINFO, and IndMED for randomized controlled trials (RCTs) of yoga for patients with depressive disorders and individuals with elevated levels of depression were included. Twelve RCTs with 619 participants were included. Despite methodological drawbacks of the included studies, yoga could be considered an ancillary treatment option for patients with depressive disorders and individuals with elevated levels of depression~\cite{cramer2013yoga}.
A similar study was conducted by the team of authors on yoga for anxiety. Eight RCTs with 319 participants (mean age: 30.0–38.5 years) were included. They conclude yoga might be an effective and safe intervention for individuals with elevated levels of anxiety~\cite{cramer2018yoga}. 
Since software engineers have a comparatively high likelihood to develop anxiety and/or depression disorders over the course of their career~\cite{ostberg2020methodology}, this evidence is of much interest for the study at hand.

To compare in between different yoga styles of practice, Cowen et al.~\cite{cowen2005physical} had twenty-six healthy adults age 20–58 (Mean 31.8) participated in six weeks of either astanga yoga or hatha yoga class. Significant improvements at follow-up were noted for all participants in diastolic blood pressure, upper body and trunk dynamic muscular strength and endurance, flexibility, perceived stress, and health perception~\cite{cowen2005physical}. The improvements differed for each group when compared to baseline assessments. The astanga yoga group had decreased diastolic blood pressure and perceived stress, and increased upper body and trunk dynamic muscular strength and endurance, flexibility, and health perception. Improvements for the hatha yoga group were significant only for trunk dynamic muscular strength and endurance, and flexibility. The findings suggest that the fitness benefits of yoga practice differ by style~\cite{cowen2005physical}.
The next section hence details the benefits specifically evidenced in Hatha yoga, which is the style practiced in our intervention.

%In educational environments, Khalsa et al.~\cite{khalsa2012evaluation} evaluated potential mental health benefits of yoga for adolescents in secondary school. Students were randomly assigned to either regular physical education classes or to 11 weeks of yoga sessions based upon the Yoga Ed program over a single semester. Students completed baseline and end-program self-report measures of mood, anxiety, perceived stress, resilience, and other mental health variables. Independent evaluation of individual outcome measures revealed that yoga participants showed statistically significant differences over time relative to controls on measures of anger control and fatigue/inertia. Most outcome measures exhibited a pattern of worsening in the control group over time, whereas changes in the yoga group over time were either minimal or showed slight improvements. These preliminary results suggest that implementation of yoga is acceptable and feasible in a secondary school setting and has the potential of playing a protective or preventive role in maintaining mental health~\cite{khalsa2012evaluation}.

\subsection{The Effectiveness of Hatha Yoga}
For specifically Hatha Yoga, there are two meta analysis studies.

Hofmann~\cite{hofmann2016effect} carried out a meta analysis that identified 17 studies (11 waitlist controlled trials) totalling 501 participants who received Hatha yoga and who reported their levels of anxiety before and after the practice and found them reduced~\cite{hofmann2016effect}.
Furthermore, Huang et al.~\cite{huang2013effects} implemented a quasiexperimental design with recruited 63 female community residents in New Taipei City aged 40–60 years where Perceived Stress Scale revealed significantly lower scores after practice~\cite{huang2013effects}.
Again, due to the often high levels of stress experienced by software engineers, these studies promise Hatha yoga as beneficial intervention.

Luu et al.~\cite{luu2016hatha} searched MEDLINE, Scopus, and PsycINFO databases for experimental studies testing the effects of Hatha yoga (acute bouts, short-term interventions, longer-term interventions) on executive function (EF). A total of 11 published studies revealed that Hatha yoga shows promise of benefit for the EF in healthy adults, children, adolescents, healthy older adults, impulsive prisoners, and medical populations (with the exception of multiple sclerosis)~\cite{luu2016hatha}.
Given the complexity of cognitive tasks that software engineers carry out, the benefits to the executive function are strongly supporting the choice of Hatha yoga as well-being intervention.

\subsection{Consequences for job performance and outcomes proposed}
To assess the evidence regarding the effectiveness of yoga programmes at work, Puerto et al. identified 1343 papers, where 13 studies met the inclusion criteria. Nine out of 13 trials were classified as having an unclear risk of bias. The overall effects of yoga on mental health outcomes were beneficial, mainly on stress. The findings of this study suggest that yoga has a positive effect on health in the workplace, particularly in reducing stress, and no negative effects were reported in any of the randomized controlled trials~\cite{puerto2019yoga}.

The dissertation by Daane~\cite{daane2018yoga} investigated yoga as a means of increasing job satisfaction in the workplace. Her sample of 32 yoga students was surveyed on yoga practice, exercise habits, past yoga experience, and levels of job satisfaction. It was predicted that students who had practiced yoga would have increased levels of job satisfaction. Results of an independent samples t test did not support the proposed hypothesis. Similar to this study, our results did not reveal an improvement in the participants personal well-being.

\subsection{Moderating Factors and Conditions for Mindfulness}
%Which type of meditation to use and necessary conditions considered.
Mindfulness meditation can be an on-the-spot intervention in workplace situations~\cite{hafenbrack2017mindfulness}.
Hafenbrack identifies three necessary conditions for an on-the-spot mindfulness intervention to be effectively used: Employees must be aware that they are in a problem situation, they must be aware of on-the-spot mindfulness intervention as an available tool, and they must actually engage in the meditation. 
Hafenbrack also describes the limitations of such engagement: It is possible that some people gain less benefit from meditation than others, e.g., defensive pessimists disproportionately harness anxiety to motivate themselves to prepare for future challenges. On-the-spot mindfulness meditation may thus have more detrimental effects on their performance than for individuals who do not employ that strategy. There are also differences across national cultures in how people conceptualize time and the ways in which they are judgmental towards others. These factors may moderate the relationships between different forms of mindfulness and various outcomes~\cite{hafenbrack2017mindfulness}.

\subsection{Well-being Interventions in SE}

Among the few well-being interventions in software engineering (SE) based on mindfulness practices, findings have shown positive outcomes. Penzenstadler et al.~\cite{penzenstadler2022take} ran a series of breathwork interventions with computer workers and found their well-being increasing over the course of the intervention, both qualitatively and quantitatively. Similar to that intervention, we used pre and post surveys and complemented with qualitative data.

Heijer et al.~\cite {den2017don} studied the impact of mindfulness on agile software teams in a two-month intervention, where mindfulness was practised for three minutes during stand-up meetings with 60+ participants from eight companies. The findings showed improved perceived effectiveness, decision-making, and listening. However, a limitation was the use of non-standard questionnaires. Additionally, Bernardez et al.~\cite{bernardez2023empirical,bernardez2020effects}  conducted a series of studies on mindfulness for software engineers showing that these interventions have positive effects on their mental well-being and self perception.

\iffalse
In that line, Romano et al.~\cite{romano2024MOOD} most recently proposed a mindfulness-based stress reduction (MBSR) platform implementation called Bravo, an intervention based on decades of MBSR research by Kabat-Zinn~\cite{kabat2003mindfulness}. % do we have a PDF?
%I do not agree to include citations that do not add value to our background. Plus that short paper does not say that it is based on Kabat's work. - BP: Ok. Agreed. I thought you wanted to include it because it's most recent and you had wanted them to quote us (hence my conclusion was you find it relevant the other way around).
\fi

In the article at hand, we present the \textbf{first study on using the modality of physical yoga poses, called yoga asana, with a software engineering population}.

\section{Methodology} \label{sec:Methodology}

This section explains the design, data collection and data analysis of our study. Additionally, it also elaborates on more methodological details.

\subsection{Research Design}
This study followed the quasi-experiment mixed-method design since our main goal was to explore whether the yoga intervention positively impacted software engineers' general well-being (measured by psychometric instruments). Based on Maciejewski \cite{maciejewski2020quasi}, quasi-experiments are observational studies where participants self-select to be included in an intervention (lack of randomisation), and there is a lack of control group. For this research, our participants were recruited by invitation from one company; however, they decided if they wanted to participate. Additionally, our (initial) control group was formed through participant self-selection. Further, our study employed a mixed-method approach, gathering qualitative and quantitative data.

\subsection{Intervention}

The intervention started with the invitation to participate in the programme. Later, participants received a link to the entry survey (including the informed consent). 
The programme lasted eight weeks. Participants had a 45-minute Hatha yoga session every Wednesday from 8:00 AM to 8:45 AM imparted by an experienced Yoga instructor. These sessions focused on the principles of Hatha yoga (5 min), incorporating physical postures (30 min), breathing exercises (5 min), and relaxation techniques (5 min).

\begin{figure}
    \centering
    \includegraphics[width=1\linewidth]{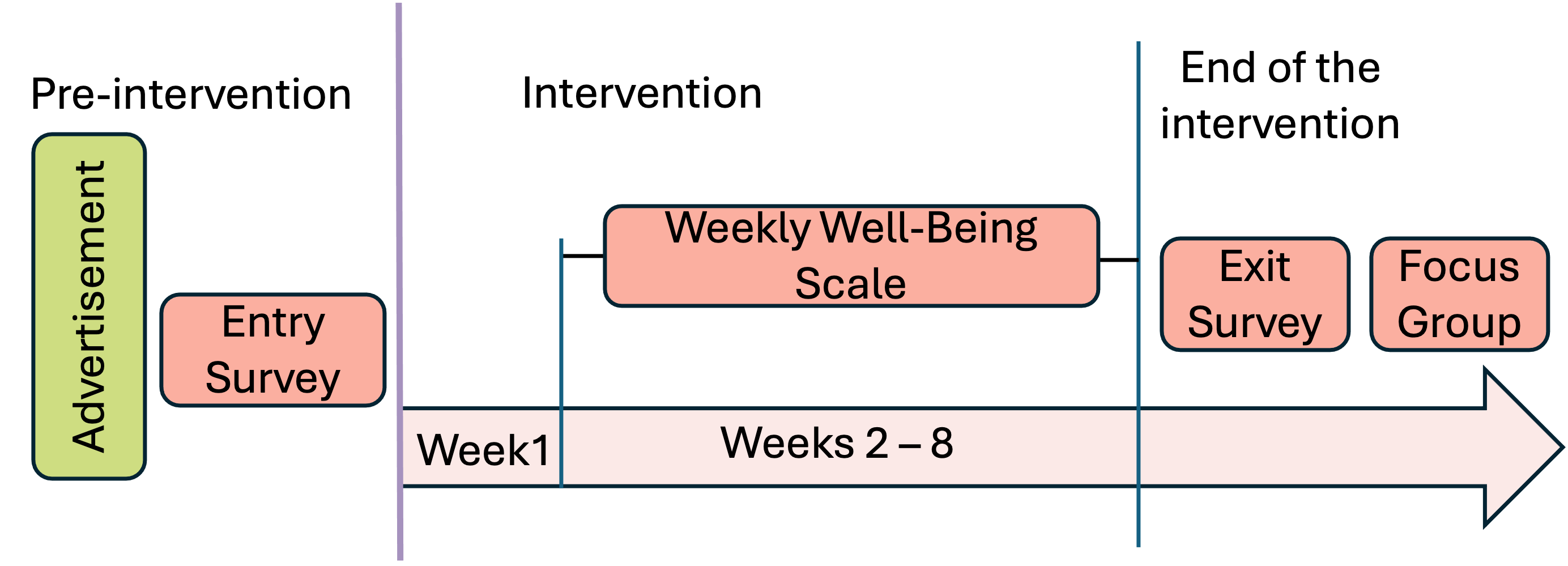}
    \caption{General Methodology of the Study}
    \label{fig:intervention}
\end{figure}

Additionally, participants receive a weekly reminder and link to fill in a weekly well-being scale to measure their well-being. 
After the eight weeks, participants completed an exit survey. When the intervention concluded, we invited participants to give an interview. However, due to the lack of positive answers, we decided to have a focus group with the organisers who were also participants in the intervention.
Figure \ref{fig:intervention} is the visual representation of the intervention.

\subsection{Company, Population and Inclusion Criteria}

The intervention occurred in an AI and software company dedicated to developing a comprehensive software stack for autonomous driving and advanced driver-assistance systems. The company has 501–1,000 employees, and its culture appears to be people-centred, with a strong emphasis on values-driven behaviour. Our target was software developers in general, so the advertisement was sent out to everyone in the company. The call was shared in the company's Slack space, and there were posters with the invitation in the elevators and shared coffee kitchen spaces. 

\subsection{Data Collection}
This section describes the three data collection strategies we used to gather data. See Figure \ref{fig:intervention} to visualise the flow and organisation of our data collection process.

\subsubsection{Entry and Exit Survey}

We tailored an entry and exit survey composed of six psychometric instruments. 

The first part of the survey asked participants to choose their alias and if they already have a personal well-being practice.

% Later, the psychometric instruments came in the following order: The Schutte Self Report Emotional Intelligence Test (SSEIT), Resilience (RS-14), Self Regulation (Short Form Self-Regulation Questionnaire (SSRQ)), Self Transcendence (Self-Transcendence Scale (STS)), Self-perceived Success (The Flourishing Scale) and Brief Resilience Coping Scale (BRCS).

Later came the psychometric instruments that integrate the survey. We considered several areas that compound individual well-being to get a complete picture. Those areas were emotional well-being, which refers to understanding and managing feelings (SSEIT). Resilience (RS-14) since it is a significant psychological predictor of well-being~\cite{izydorczyk2019resilience}. Coping strategies (BRCS) strongly relate to positive physical and psychological health outcomes in stressful circumstances~\cite{taylor2007coping}, leading to better long-term well-being. Self-perceived Success (The Flourishing Scale) measures an individual’s self-perceived success and optimal functioning across all areas of life, reflecting the core elements of overall well-being~\cite{logan2023vitality}. Self-regulation (SSRQ) since higher self-regulation is linked to greater psychological well-being, including growth, purpose, relationships, and self-acceptance~\cite{hofer2011self}. 
Finally, Self Transcendence (Self-Transcendence Scale (STS)) to measure the ability to derive a sense of well-being through
cognitive, creative, social, spiritual, and introspective avenues~\cite[p.~1]{reed2018self}.

The \textbf{The Schutte Self Report Emotional Intelligence Test (SSEIT)} \cite{schutte1998development} is an instrument to measure emotional intelligence developed by Dr. Nicola Schutte and her colleagues in 1998. The authors used the model of emotional intelligence of Salovey and Mayer as the conceptual foundation for the items used in the scale. It contains 33 items and uses a five-point Likert scale going from ``strongly disagree" to ``strongly agree".
 
The \textbf{14-Item Resilience Scale (RS-14)} was developed by Wagnild \cite{wagnild2011resilience} as a shorter version of the original 25-item RS. This instrument measures five characteristics of resilience, namely: meaning and purposeful life, perseverance, equanimity, self-reliance, and existential aloneness \cite{aiena2015measuring}. It uses a 7-point Likert-type response format and is widely used in different fields.

\textbf{Short Form Self-Regulation Questionnaire (SSRQ)}~\cite{carey2004psychometric} contains 31 items. It is the short version derived from the Self-Regulation Questionnaire (SRQ) \cite{brown1999self} that was designed to measure self-regulation capacity across seven processes. Responses are rated on a 1–5 scale (strongly disagree to strongly agree) and can be summed up to generate a total score.

\textbf{Self-Transcendence Scale (STS)}, ``Self-transcendence" (ST) refers to the ability to broaden personal boundaries and focus on perspectives, activities, and goals beyond oneself, while still recognising the value of the self and the present context\cite{haugan2012self}. ST can result in personal transformation, enhancing well-being and improving quality of life \cite{teixeira2008self}. The STS was developed by Reed in 1986 and contains 15 items that address specific behaviours or perspectives associated with expanding self-boundaries in various ways. It includes inward expansion through introspective activities, outward expansion through interactions with others, and temporal expansion by living in the present or adopting perspectives on the past and future that enrich the present\cite{haugan2012self}.

The \textbf{Flourishing Scale (FS)} \cite{diener2009new} is a concise 8-item measure that assesses the respondent's self-perceived success in key areas like relationships, self-esteem, purpose, and optimism. It yields a single score representing psychological well-being.

\textbf{Brief Resilient Coping Scale (BRCS)} \cite{sinclair2004development} is a 4-item measure specifically designed to assess an individual's tendency to cope with stress in highly adaptive ways. Each item in this brief questionnaire targets a different aspect of adaptive coping strategies, encouraging respondents to reflect on how they manage stress in various situations

\subsubsection{Weekly Well-being Scale}
Every week participants answered the weekly mini-survey including the World Health Organisation-Five Well-Being Index (WHO-5) and an open question at the end for participants to elaborate on their week if they wanted to.
The WHO-5 is a brief self-reported assessment of current mental well-being using five questions, these questions are answered with a six-point scale from ``All of the time" to ``At no time". Table \ref{tab:WHO-5} shows the questions of the weekly scale.

\begin{table}[h]  
\caption{WHO-5 Well-being Index}   
\centering   
\begin{tabular}{p{1cm} p{7cm}}  
\hline\hline   
 \textbf{No.} &  \textbf{Questions}
\\ 
\hline   
 WHO-1 &  I have felt cheerful in good spirits. \\[1ex]  
 WHO-2 &  I have felt calm and relaxed.  \\[1ex]
 WHO-3 &  I have felt active and vigorous.\\[1ex]
 WHO-4 &  I woke up feeling fresh and rested. \\[1ex]
 WHO-5 & My daily life has been filled with things that interest me.  \\[1ex]
Open q. & Is there anything else you'd like me to now? \\[1ex]
\hline  
\end{tabular}  
\label{tab:WHO-5}
\end{table}  

\subsubsection{Focus Group}

We conducted a focus group with the company's intervention coordinators to better understand the internal experts' individual experiences and evaluate the intervention. The three participants were in managerial positions and were in charge of logistics within the company. We asked them to answer the questions from two perspectives, as participants and 
organisers of the company's intervention. Table \ref{tab:questions} shows the questions we used as interview guide.

\begin{table}[h]  
\caption{Focus group questions}   
\centering   
\begin{tabular}{p{1cm} p{7cm}}  
\hline\hline   
 \textbf{No.} &  \textbf{Questions}
\\ 
\hline   
 1 &  What was your personal experience of the course? \\[1ex]  
 2 &  What is your impression of the overall group experience?  \\[1ex]
 3 &  How does your experience in this intervention compare to other well-being practices that you do?\\[1ex]
 4 &  Within your company, what other well-being practices have you offered in the past, and how do you think they compare to this intervention? \\[1ex]
 5 &  What would you personally wish the next well-being intervention to look like? \\[1ex]
 6 &  What do you think the potential pool of participants will wish for?  \\[1ex]
\hline  
\end{tabular}  
\label{tab:questions}
\end{table}

\subsection{Data Analysis}

This section explains how the qualitative and quantitative data were analysed.

\subsubsection{Statistical Analysis of Instruments}
Data analysis of the psychometric scales was conducted using RStudio. After cleaning the database, we obtained descriptive statistics (mean and standard deviation) and analytic statistics (normality tests and independent samples t-test). We considered the significance level of 0.05 (P = 0.05) for all statistical tests. To compare the entry and exit surveys, we initially chose the independent samples t-test since our groups had different numbers of participants due to dropouts. Additionally, we performed a paired t-test using data from participants who completed both the entry and exit surveys. This allowed us to account for within-subject differences and maximise the statistical power for this subset of participants despite the smaller sample size. We included this analysis to understand better changes among those who fully participated in the intervention. Further, since our control group was very small and became even smaller by the end of the intervention, we decided not to include it in any statistical tests, as the statistical power was already compromised. See our repository \cite{anonymous_2025_14721592} for the database and code. 

For the weekly scale, we only report the means per week. We calculated the scores by averaging the responses of all participants who completed the survey each week.

\subsubsection{Qualitative Analysis of Focus Group}

%how to create themes in nvivo   https://youtu.be/zJNpGK3HBgI

To analyse the data gathered from the focus group we followed the guidelines of thematic analysis by Braun and Clarke \cite{clarke2021thematic}. We decided to perform it inductively, that is, codes and themes were derived directly from the data. The first and second authors went through the transcripts to become familiar with the data, as stated in the first step. Then the initial codes were generated, compared and discussed to reach agreement on their interpretation and to ensure consistency in the coding process. Later, the themes were identified, reviewed and defined to finally write up the results.

\subsubsection{Reflexivity}

The first author has a bachelor's degree in psychology and a master's degree in social work, and brings a deep understanding of human behaviour and social dynamics to the study. Her background and expertise in psychometrics equip her with the skills to explore the psychological aspects of well-being, such as stress management, coping strategies, and interpersonal relationships, which are crucial in the context of software engineering work environments.

Conversely, the second author, who holds a PhD in software engineering, offers expertise in the technical aspects of software development and extensive education as a yoga teacher. Their knowledge can shed light on the work-related factors that impact well-being, such as workload, project deadlines, and the use of technology in the workplace. 

The mix of backgrounds and approaches allows to critically evaluate interventions that address stress and well-being in the software engineering field

\subsection{Ethical Considerations}

This study adhered to the ethical research guidelines recommended by our university. Additionally, the study received approval from the country's ethics agency. %Swedish Etikprövningsmyndigheten\footnote{https://etikprovningsmyndigheten.se}. 
All participants gave their informed consent.

Participants were comprehensively briefed on the study's objectives, methods, and potential risks. They were also informed of their right to withdraw from the study at any time without any consequences.

To ensure participants' privacy, all personal identifying information was kept strictly confidential. All collected data, including transcripts and audio recordings, was anonymised and securely stored.

\section{Results} \label{sec:Results}

In this section, we report the results of the quantitative and qualitative data. 

The intervention started with twenty-nine participants filling in the entry survey and finished with fourteen exit survey responses for the intervention group. For the control group, seven participants filled in the entry survey and five the exit survey. 

The intervention group and the control group had similar demographics: We had a balance in terms of gender ~50/50 men/women (no one identified as non-binary). All participants were at at advanced stages of their career with ~10-15+ years experience. They all held a university education (either MSc or PhD), and we had about 33\% in leadership roles (program manager, project manager, product manager, engineering manager) and roughly 66\% engineers. These percentages are quoted as "roughly" since some participants have overlapping functions and do not qualify as strictly one or the other.
About 90\% of participants were in technical roles in engineering and about 10\% in human resources, communication and business management.
Of the control group, 80\% were in technical roles and about 20\% in human resources, communication or business management.

\subsection{Quantitative Analysis}

\begin{figure}
    \centering
    \includegraphics[width=0.8\linewidth]{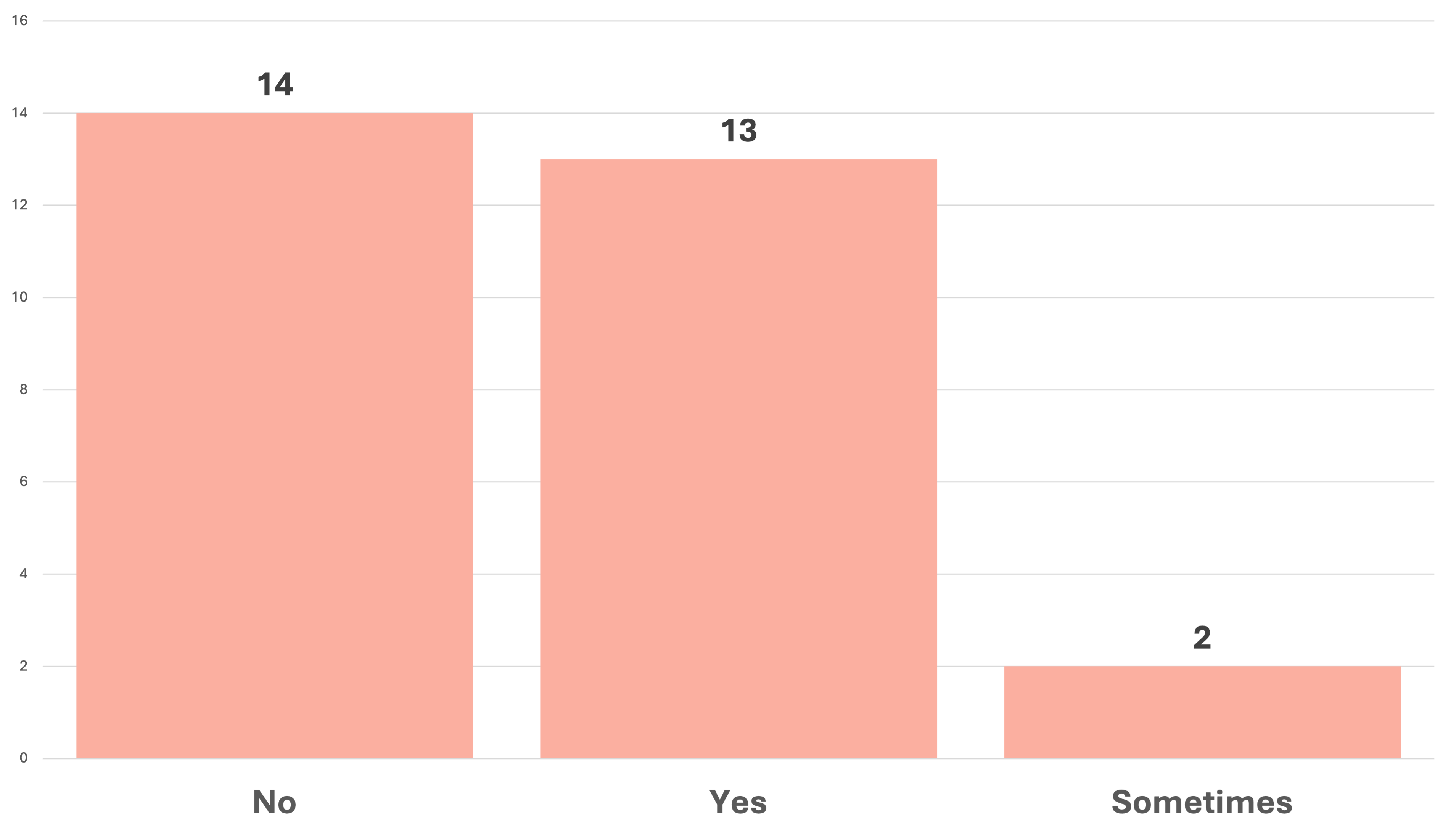}
    \caption{Participants' Having Well-being Practices Before the Intervention}
    \label{fig:practice}
\end{figure}

Answers to the question about participants currently having a well-being practice are shown in Figure \ref{fig:practice}. The majority (14) answered with a ``No", meanwhile 13 participants said they have a practice and 2 participants that only sometimes. 

\begin{figure}
    \centering
    \includegraphics[width=0.8\linewidth]{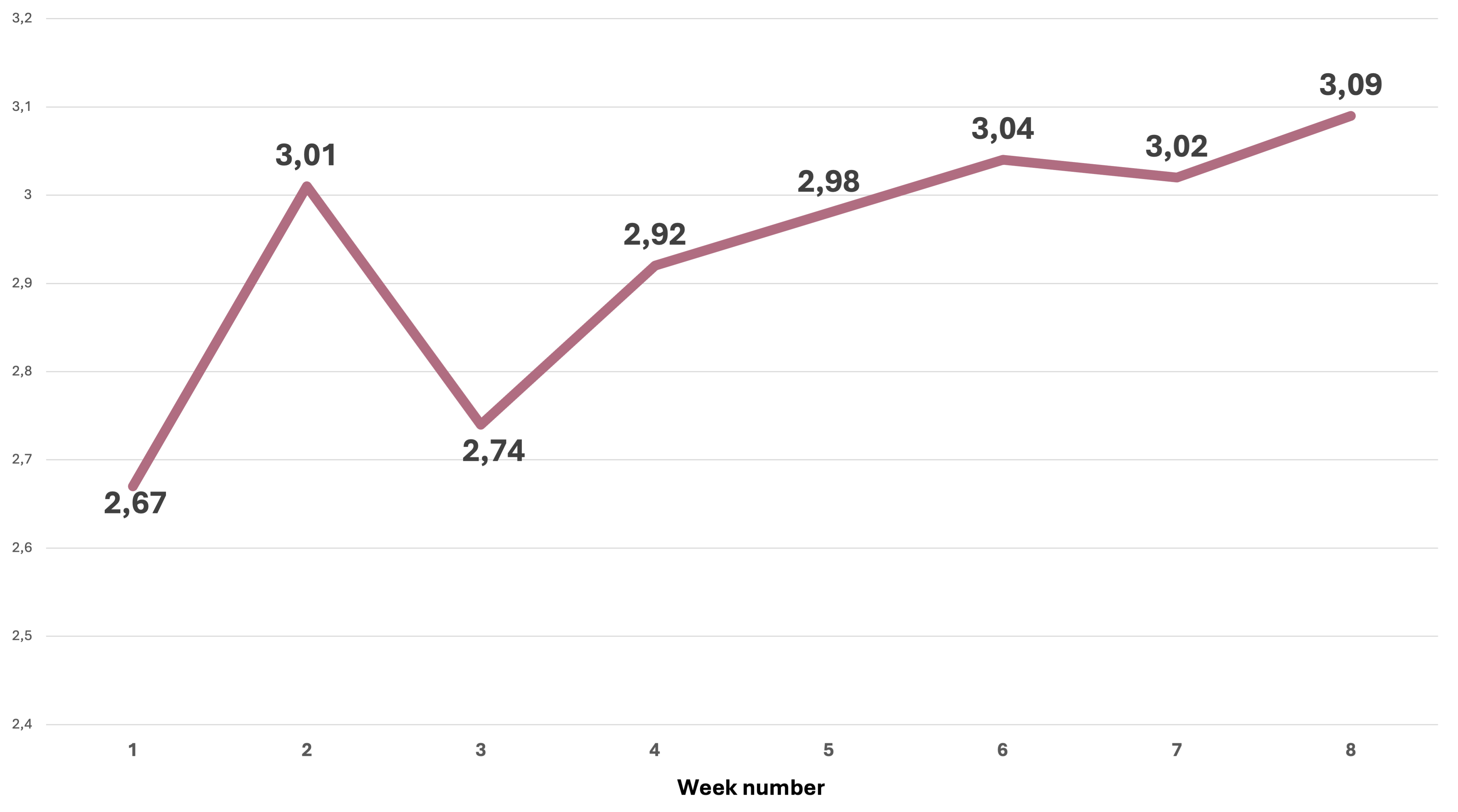}
    \caption{Participants' Weekly Well-being Score. We considered all participants means that answered each week.}
    \label{fig:tunein}
\end{figure}

The results of the weekly scale are shown in Figure \ref{fig:tunein}. It is visible that overall, participants had a higher level of general well-being at the end of the intervention compared to the initial one in week 1.

Regarding the psychometric instruments, we first performed the Shapiro test to assess the normality. Table \ref{tab:shapiro_test_combined} shows the results of the normality test for each psychometric instrument (W). The p-values are shown in brackets, all of which are greater than 0.05. Therefore, we do not reject the null hypothesis, indicating that the data can be assumed to follow a normal distribution.

\begin{table*}[h]
\centering
\caption{Shapiro-Wilk Normality Test Results}
\label{tab:shapiro_test_combined}
\begin{tabular}{l l c c c c c c}
\hline
\textbf{Test} & \textbf{Group} & \textbf{SSEIT} & \textbf{Resilience} & \textbf{SelfRegulation} & \textbf{SelfTransformation} & \textbf{SelfSuccess} & \textbf{Coping} \\ 
\hline
\multirow{2}{*}{Pre-test} & \textbf{Int En} & 0.982 (0.886) & 0.966 (0.459) & 0.969 (0.540) & 0.951 (0.192) & 0.972 (0.623) & 0.942 (0.115) \\ 
                          & \textbf{Cont En} & 0.780 (0.026) & 0.937 (0.610) & 0.926 (0.517) & 0.967 (0.876) & 0.900 (0.332) & 0.915 (0.432) \\ 
\hline
\multirow{2}{*}{Post-test} & \textbf{Int Ex} & 0.971 (0.895) & 0.908 (0.147) & 0.970 (0.875) & 0.975 (0.933) & 0.935 (0.356) & 0.963 (0.764) \\ 
                           & \textbf{Cont Ex} & 0.964 (0.838) & 0.908 (0.453) & 0.764 (0.040) & 0.804 (0.087) & 0.813 (0.104) & 0.964 (0.833) \\ 
\hline
\end{tabular}
\end{table*}

\begin{figure*}
    \centering
    \includegraphics[width=0.8\linewidth]{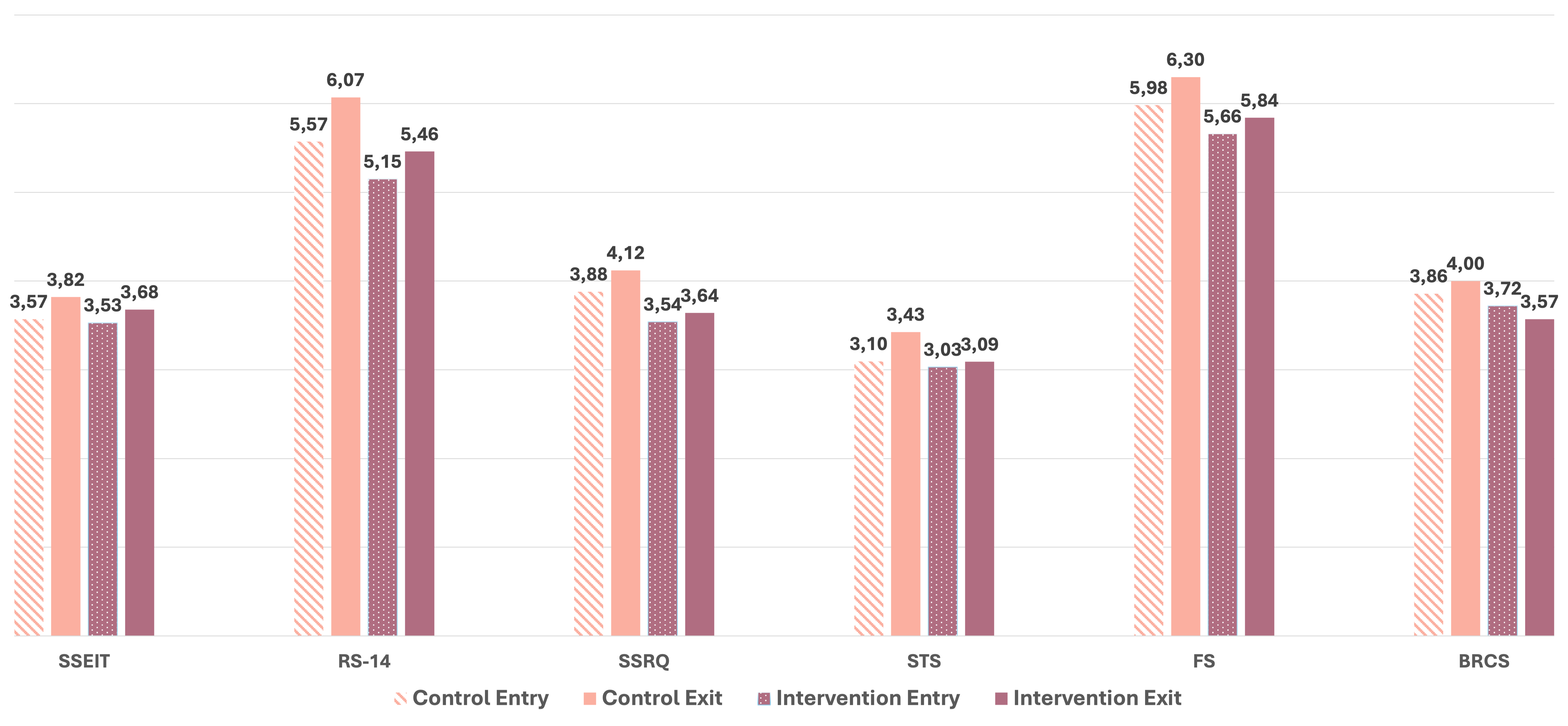}
    \caption{Means Comparison Between the Intervention (pre and post-programme) and Control Group}
    \label{fig:expGroup}
\end{figure*}

We then calculated the descriptive statistics for each scale. Figure \ref{fig:expGroup} shows the mean scores for each psychometric instrument per group and the visual comparison of all the means. The differences between the entry and exit surveys and the control group are minimal. Based only on the means, the control group showed a better improvement in all scales in comparison to the intervention group. The difference in the means of the intervention group was slightly higher and even one scale (STS) had a decrease after the intervention.

To explore the differences between pre and post-intervention, we initially performed an independent t-test; the results are presented in table \ref{tab:ind_t_test_results}. There were no significant differences in any scale after the intervention finished. Then, we also performed a paired t-test with only the 14 participants who completed pre- and post-intervention surveys to gain additional insight into the data. Table \ref{tab:t_test_results} presents the results. Although the overall findings remain quantitatively non-significant, the paired t-test provided a clearer view of the data for participants who fully engaged in the intervention.

\begin{table}[h]
\centering
\caption{Results of the Independent T-Tests for Psychometric Scales}
\label{tab:ind_t_test_results}
\begin{tabular}{l c c c}
\hline
\textbf{Scale} & \textbf{t value} & \textbf{df} & \textbf{p value} \\ 
\hline
Emotional Intelligence & -1.123 & 21.389 & 0.2739 \\ 
Resilience & -1.2905 & 23.711 & 0.2093 \\ 
Self-Regulation & -0.6949 & 21.601 & 0.4945 \\ 
Self-Transcendence & -0.4783 & 22.945 & 0.6370 \\ 
Self-Perceived Success & -0.9278 & 28.728 & 0.3612 \\ 
Coping & 0.7535 & 20.693 & 0.4597 \\ 
\hline
\end{tabular}
\end{table}

\begin{table}[h]
\centering
\caption{Paired T-test Results for Each Scale}
\label{tab:t_test_results}
\begin{tabular}{l c c c}
\hline
\textbf{Scale} & \textbf{t value} & \textbf{df} & \textbf{p value} \\ 
\hline
Emotional Intelligence & -0.75378 & 13 & 0.4644 \\ 
Resilience & -0.18751 & 13 & 0.8542 \\ 
Self-Regulation & -0.65387 & 13 & 0.5246 \\ 
Self-Transcendence & -0.08407 & 13 & 0.9343 \\ 
Self-Perceived Success & -0.30439 & 13 & 0.7657 \\ 
Coping & 0.25320 & 13 & 0.8041 \\ 
\hline
\end{tabular}
\end{table}

\subsection{Thematic Analysis}

\begin{figure}
    \centering
    \includegraphics[width=1\linewidth]{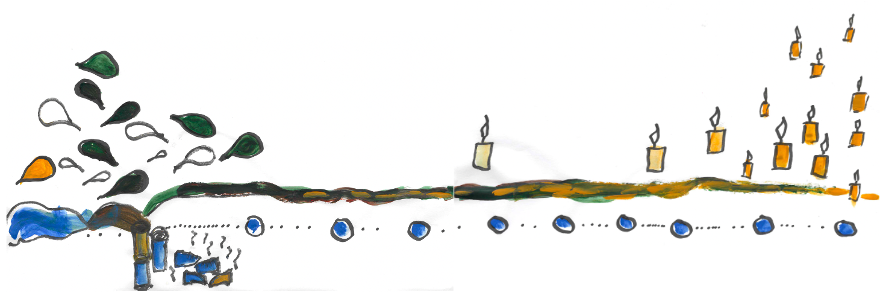}
    \caption{Participant's Symbolic Journey During the Yoga Programme}
    \label{fig:focusfig}
\end{figure}

\begin{figure}
    \centering
    \includegraphics[width=1\linewidth]{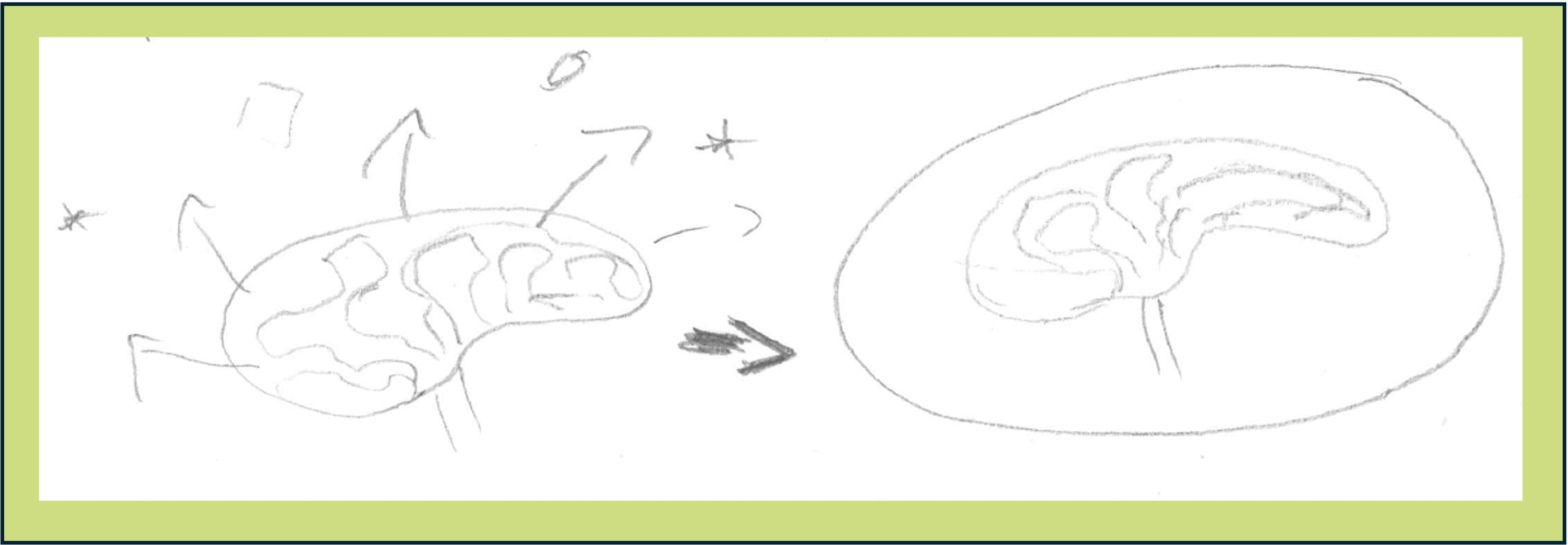}
    \caption{Two Brains [``One representing a wandering mind, and the other, a focused mind. For me, this captures the essence of the journey"]. Image From Focus Group}
    \label{fig:focusfig2}
\end{figure}

Three themes were generated from the focus group data analysis and are described below. Figures \ref{fig:focusfig} and \ref{fig:focusfig2} are representations of the focus group's participant experience during the yoga intervention.

\subsubsection{Theme 1: Individual Benefits and Shared Reflections in Practice}

This theme describes the impact of the intervention on personal and group levels. We identified three sub-themes that show how yoga influenced participants' well-being, fostered group dynamics, and evoked symbolic representations of the practice. 

\textbf{Sub-theme 1: Personal Benefits.}
This sub-theme focuses on the individual gains participants experienced from the yoga sessions, spanning physical, mental, and emotional well-being. Participants commented how the intervention helped them manage stress and enhance emotional balance, emphasising how breathing techniques contributed to relaxation and focus. 

This participant explained how yoga offered them more than physical or mental benefits. The quote below shows that yoga helped them relieve stress, promote overall well-being, and contribute to their professional life by enhancing their cognitive abilities. Specifically, yoga improved mental clarity, focus, and knowledge acquisition, which helped them perform better in their work. 

 \interviewquote{So I want to say that it's not only yoga and well-being. It's stress relief, but it's also cognitive input to my professional work life that helps me… it adds value to other things than only to my body and mind but also to my cognition, my knowledge.}  

Participants shared stories of overcoming initial hesitation towards yoga, with some noting their previous negative experiences with fast-paced classes. In contrast, this intervention's structured and mindful pace was described as relaxing and immediately impactful, encouraging participants to remain open to future yoga sessions. One participant, initially sceptical of yoga, reflected on how they overcame the barrier of waking up early to attend the sessions and found the practice deeply relaxing. These narratives explain how yoga fostered a sense of mindfulness beyond physical benefits, enabling participants to separate rational thought from emotional stress.

\textbf{Sub-theme 2: Group Experience.}
Yoga also had a significant effect on the collective experience of participants. The organisers commented that a core group of about seven participants consistently attended the sessions and provided highly positive evaluations of the practice. While there was a drop in participation after the first two sessions, attendance stabilised, and those who continued to attend reported looking forward to the classes and appreciating their effects.

 \interviewquote{I can remember a few times where someone actually either wrote on Slack or came to me saying something like: “I felt really bad in the morning and after yoga, I felt so much more ready for the day in a positive mindset”.}  

Participants gave the organisers generally good feedback, and this was shown in practice when they returned to class after missing a week and even joined online due to difficulties in commuting. The yoga sessions were beneficial on an individual level and created a shared space for relaxation within the company. One example is the ``words of wisdom" (as commented by one participants) shared during the classes, which were described as having a lasting impact, with participants feeling empowered to pass on these lessons to others outside of the sessions.

\textbf{Sub-theme 3: Visual and Symbolic Representations.}
A unique aspect of the participants' experience was how they described yoga through visual and symbolic representations. Participants used imagery to capture the mental and emotional states fostered by the sessions. For example, the colour blue (in an image done in the focus group) was repeatedly mentioned, symbolising peace and harmony, with one participant visualising blue bubbles during breathing exercises to represent a sense of calm. 

 \interviewquote{Then also my peace during the sessions became better. So that represents the blue dots, all the sessions we've had and that they were really like harmonised and peaceful.} 

Conversely, darker colours were used to depict confusion or unclear mental states early in practice, which gradually transitioned to lighter colours, symbolising clarity and calm as the sessions progressed.
Other visual metaphors included two brains, one representing a wandering, distracted mind and the other symbolising the focused state achieved through yoga. Participants also highlighted the symbolism of yoga mats, which sparked discussions around them. The candles used in the sessions were described as a symbol of tranquillity, contrasting with chaotic external conditions, such as the inconvenience of practising near smelly shoes (week 1 due to the small room capacity).
These visual and symbolic representations reflect participants' deep mental and emotional engagement with the practice.\\

\subsubsection{Theme 2: Organisational Support and Logistical Challenges in Implementing the Programme}

This theme explains the relationship between organisational support and the logistical challenges of implementing the yoga programme. Organisers acknowledged the company's commitment to promoting well-being, recognising its role in encouraging employee engagement. However, they also highlighted limitations within the organisation that could hinder participation. 

Logistical factors, including room characteristics, scheduling preferences, and resource availability, significantly influenced participants' experiences. Additionally, the complexities of securing approval and coordinating sessions illustrated the challenges organisers faced, particularly when balancing their dual roles as both organisers and participants. Overall, this theme emphasises the need for ongoing organisational support and effective logistical planning to create an inclusive environment that encourages participation in well-being initiatives.

\textbf{Sub-theme 1: Company's Role in Supporting Well-Being Initiatives.}
Participants commented on the company's role in supporting the yoga sessions as part of its broader well-being initiatives. The intervention was seen as an opportunity for the company to demonstrate its commitment to employee health, and several participants expressed high appreciation for the company's involvement in promoting well-being practices. Providing such interventions within the workplace was viewed favourably, with many recognising that workplace-based yoga sessions offered logistical advantages compared to external options like gym memberships or yoga studio subscriptions.

 \interviewquote{Once a year we have wellness day, where we get presentations by different companies for like advertisements on well-being and what to do.} 

Despite this, there was also discussion about the limitations of the company's well-being initiatives. While the yoga sessions were well-received, participants noted that other employees' priorities or workloads might interfere with fully engaging in such programs. 

 \interviewquote{Some have meetings at 8:30 And then some have meetings at 9. There's always someone who has the next important meeting. It's so hard to fit everyone's discussion.} 

This points to the need for well-being interventions that not only exist but are integrated into a broader culture of health within the organisation, encouraging participation from a wider range of employees.

\textbf{Sub-theme 2: Logistical and Environmental Factors.}
Logistics and the physical environment played a significant role in shaping participants' experiences with the yoga sessions. Participants described how environmental features like using (fake) candles helped create a calming atmosphere conducive to yoga. However, challenges related to space and resources were also mentioned, such as the availability of yoga mats and differing preferences for class locations.

 \interviewquote{And then there was a huge discussion where it should be. Should it be in  [room's name]? which is a big room that we have upstairs. Or should be downstairs in a smaller room?} 

Furthermore, the hybrid option—allowing employees to participate either in person or online—was seen as a valuable addition, especially for those working from home, and it contributed to the overall accessibility of the program. 

\textbf{Sub-theme 3: Challenges and Efforts in Organising Well-Being Programs.}
Participants reflected on the challenges involved in the general planning, such as getting approval and securing funding for the sessions, with one describing the process as a ``chaotic journey" that required considerable effort to bring everything into place. 

 \interviewquote{This was chaos for organising. And until everything fell into place, which is along the journey, it took quite a lot of effort.}

There was initial resistance from the company, and it took time to convince key decision-makers of the value of the intervention. The level of commitment required from the organisers was also emphasised, with organisers playing dual roles—coordinating and participating in the sessions. This dual role added a layer of complexity, as the line between organiser and participant blurred. Participants noted that a significant amount of work went into ensuring the program ran smoothly, from logistical planning to recruitment and retention efforts. The organisers' high dedication was necessary to overcome these obstacles and implement the program effectively.

\subsubsection{Theme 3: Perception and General Feedback}

This theme focuses on employees' perceptions of, feedback on, and responses to the yoga programme in the company. It shows a landscape of interest, engagement, and organisational context. Participants shared varying perceptions of the company's efforts to promote yoga and well-being initiatives. They wanted more information about future courses and a greater understanding of yoga practices. There was a noticeable awareness of stress levels within the company, motivating employees to seek additional well-being strategies. 

 \interviewquote{Not everyone but a lot needed it. Because we are stressed and consciously, we don't admit this.}

However, while some employees were enthusiastic and inquired about upcoming sessions, there was a recognition that participant commitment might not always match the organisers' dedication to the program. To enhance participation, several suggestions for strategies to involve more employees emerged. Additionally, participants highlighted the significance of research in evaluating these programs, noting that understanding the impact and outcomes of yoga sessions could reinforce their value within the organisation.

\section{Discussion} \label{sec:discussion}

In this section, we interpret the results in a wider context, argue for the relevance of the study despite non-significant statistical results, and the threats to validity.

%\textcolor{magenta}{BP: from discussion with Robert - can we change framing?\\ 3 factors that contributed to no significant results: (1) control group flaked, (2) lot of stress at company because of end-of-year and milestones and company changes, (3) yoga cannot counteract increasing stress}

%I'm commenting this out to propose a re-written text
\iffalse
\subsection{Negative Results} The biggest challenge of our study was that the control group did not follow up on their promises to fill out the surveys. Hence, their data was so sparse that we could not use it for a meaningful comparison.

In addition, our findings do not align with similar interventions; for example, Montes and Penzenstadler~\cite{montes2023piloting} performed a four-week intervention implementing different well-being practices, yoga one of them. Their results reported positive changes in participants in a shorter intervention than ours. 

Looking at Figure~\ref{fig:tunein}, we notice a spike from week one to week two, which we attribute to the newness factor. Subsequently, we see a significant drop in rating in the third week which is most likely due to a major company milestone where a product was going public for the first time. The instructor recalled several participants commenting on this event and how stressed they were about it.
\fi

\subsection{Results in Context} % Good?
%My proposal

The interest in studying the effects of a yoga intervention at the workplace was based on the positive benefits in different contexts~\cite{markil2010hatha,bryant2015yoga,luu2016hatha}. We also considered Hafenbrack's~\cite{hafenbrack2017mindfulness} factors for on-the-spot intervention in the workplace. In the  research at hand, the yoga intervention to reduce stress among software developers showed no statistically significant effect across the six psychometric scales between pre- and post-test assessments. There was a \textbf{slight increase in mean scores} for each scale; however, the changes were not large enough to reach \textbf{statistical significance}. Further, the control group was not big enough to perform statistical tests with enough power and reliability. Our quantitative findings suggest that the intervention may not have produced measurable improvements in participants' emotional intelligence, resilience, self-regulation, stress transformation, perceived success, or coping abilities, at least within the time frame and structure of the study.

We list several potential explanations for these results. First, the frequency of the intervention --- one session per week—might not have been sufficient to create significant shifts in the psychometric outcomes. Other interventions with similar populations using mindfulness practices, such as meditation~\cite{bernardez2023empirical} and yoga~\cite{montes2023piloting}, had a higher frequency (four times per week and daily practices, respectively). This suggests that our \textbf{intervention's ``dosage'' may not have been enough} to induce substantial changes. We did not track participants' engagement outside the weekly sessions; if participants only stuck to the weekly sessions, they may not have experienced the full benefits of yoga. In addition, participants frequently mentioned the ``end of the year'' stress of having to finish a large number of tasks before the holidays. Hence, the \textbf{timing of the intervention} in the last 8 weeks before the winter break may not have been ideal (from a data collection point of view) since participants were likely to experience an increase of stress.

Another possible reason for the lack of significant findings could be the complexity of the software development tasks' stress. Software engineering is a high-strain job~\cite{wong2023mental} with a combination of high demands~\cite{teevan2022microsoft}, constant change~\cite{dasanayake2019impact} and technology-reliant~\cite{singer2010examination}. Furthermore, it is plausible that external stressors continued to affect participants, potentially overshadowing the benefits of the yoga intervention.

Interestingly, despite the absence of significant changes in the psychometric scales, the focus group analysis revealed positive feedback from participants. The analysis showed they felt more relaxed, better able to manage stress, and more mindful after attending the sessions, aligning with existing literature that suggests yoga can improve subjective well-being~\cite{ross2010health,cramer2018yoga,cramer2013yoga,cowen2005physical}. Looking at Figure~\ref{fig:tunein}, we notice a spike from week one to week two, which we attribute to the newness factor. Subsequently, we see a significant drop in rating in the third week which is most likely due to a \textbf{major company milestone} where a product was going public for the first time. The instructor recalled several participants commenting on this event and how stressed they were about it. There is no way to control for such external stressors or confounding factors.

Overall, at the end, the well-being of the group finished higher than at the beginning of the intervention. This discrepancy between the quantitative and qualitative data may suggest that the intervention had subjective benefits that were not fully captured by the psychometric tools. Participants may have experienced shifts in their stress perception or management that were more subtle, context-dependent, and not easily measurable by standardised scales. Similar to the Daane~\cite{daane2018yoga} study and their quantitative results. This also stresses the importance of considering quantitative and qualitative outcomes when evaluating intervention programmes.

\subsection{Importance of the Study}
Despite these results, why is this study important?
On a larger scientific scale, negative or null results become part of the bigger story about the intervention and what it targeted. By publishing negative results, we strengthen transparency and accountability in research. They help to interpret positive results that may have been obtained in related studies. They may adjust research designs and thereby increase the chances of success. Finally, the publication of null results will result in less bias in future meta-analysis studies, which could have incorrect conclusions if negative results are not included because they were never published. A less biased range of outcomes will ensure such meta-analyses are much more valuable.

In the particular case of our study, there are a number of confounding variables that were not possible to filter out and control in the sample size. We need these null results to redesign our experiment. We need access to negative and null results to guide us on the path to positive results.

The benefits of yoga may not be universally applicable or may require longer-term interventions, different formats, or complementary approaches to yield noticeable improvements in high-stress, cognitively demanding professions.
Hence, it is important to carefully tailor wellness interventions to the unique needs of their workforce. Rather than relying on one-size-fits-all approaches, organisations may need to explore other strategies or enhance yoga programs with additional resources like mental health support, ergonomic adjustments, or stress management training.

\subsection{Lessons Learned}

Despite the lack of measurable changes in psychometric scales, we identified several important lessons:

\textbf{[L1] Stress-Management Interventions Must Address Software Engineering Workflows.}
The intervention was during a period of high pressure for the company, characterised by year-end deadlines and critical project milestones. However, due to the software engineering dynamics, for example, product release cycles, Agile sprints, and incident response demands, stress is always a challenge when finding the right time to start an intervention. This context might increase the difficulty of engaging participants when work-related stress peaks. Future interventions should account for these patterns and align better with project timelines. These include integrating short, stress-relief activities during sprint breaks or conducting longer sessions in less intense project phases.

\textbf{[L2] Engagement Is Not Synonymous with Measurable Outcomes.}
Participants reported enjoying the weekly yoga sessions, but this positive reception did not translate into measurable improvements in any psychometric scale. It might suggest the need for future programmes to incorporate elements beyond enjoyment—such as tracking individual goals, providing reminders for daily practice, or connecting the intervention to broader organisational well-being strategies.

\textbf{[L3] Weekly Interventions Alone Are Insufficient in High-Stress Contexts.}
A single weekly session, while appreciated by participants, was insufficient to counteract the acute and ongoing stressors in the software engineering workplace. This limitation stresses the importance of integrating more frequent or accessible stress-management practices into daily routines. For example, teams might benefit from micro-interventions, such as five-minute breathing exercises or mindfulness breaks incorporated into stand-ups or coding sessions.

\textbf{[L4] Psychometric Scales Alone May Not Capture Software Engineering-Specific Stressors.}
The validated psychometric tools used in this study may not fully reflect the unique stress dynamics in software development, such as cognitive overload from debugging, context-switching, or tool-related frustrations. Although these scales measured general concepts related to well-being and resilience, their lack of sensitivity to domain-specific stressors and acute stress may have contributed to the lack of significant findings. In future interventions, we suggest additional metrics and qualitative methods tailored to the software engineering context.

\textbf{[L5] Participant Dropout: a Need for Flexibility and Individualisation.}
We had a notable dropout rate, suggesting that the one-size-fits-all approach may not meet the diverse needs of software engineers. Participants likely struggled to balance attendance with their demanding schedules, especially during a high-pressure work period. To minimise dropouts in future interventions, offering more flexible options—such as recorded sessions for asynchronous participation or shorter, on-demand activities—could better accommodate varying workloads and time constraints.

In general, we learned that it is essential to tailor intervention programmes to the unique demands and context of software engineering. By aligning interventions with team dynamics, cognitive workloads, and the cyclical structure of the work, organisations can create more effective and sustainable approaches to supporting employee well-being.

\subsection{Validity Threats}

\subsubsection{Internal Validity}
Several factors were considered to address internal validity. First, the intervention was voluntary, meaning random assignment to groups was impossible. As a result, self-selection bias likely occurred, as participants had a pre-existing interest in or experience with yoga.
Another challenge was controlling and measuring confounding variables, making it difficult to determine whether other factors influenced the intervention outcomes. To mitigate this, we attempted to use a control group to establish a baseline for comparison and conduct pre- and post-intervention assessments. However, the control group was ineffective, limiting its utility in the analysis. 

\subsubsection{External Validity}

 Our intervention was conducted in a realistic setting, making it applicable to similar work environments aiming for generalisability. The study's conditions were designed to be replicated across different companies and everyday situations, mimicking real-world scenarios. To facilitate this, we provided a detailed methodology and a replication package\cite{anonymous_2025_14721592} to allow for the reproduction of the study in diverse settings. While we acknowledge that the cultural context of the company may limit the generalizability to similar environments, the participants came from diverse backgrounds. This diversity within the participant pool may mitigate cultural constraints, suggesting that the findings could be relevant across various organisational settings, provided similar working conditions and organisational cultures exist.

 \subsubsection{Construct Validity}
To ensure construct validity, we considered several actions. For example, we used psychometric standardised tools to ensure the measurement of our variables was accurate. Data collection was triangulated with qualitative data from the focus groups to complement the scales. We also implemented a longitudinal follow-up during the intervention using a weekly tune-in to monitor changes over time. Further, we had experts in psychometrics and yoga interventions to review the instruments and methodology. These combined efforts strengthened the credibility of our findings and helped ensure that the constructs were accurately captured throughout the study.

 \subsubsection{Conclusion Validity}
Several challenges compromised the conclusion validity of the study. The small and ineffective control group, which was further reduced by participant dropouts, limited statistical power hindered the ability to draw reliable conclusions about the intervention's true effects. Additionally, external stressors, such as the end-of-year workload and critical company milestones, may have confounded the results, as these factors could have overshadowed any potential benefits of the yoga intervention. Finally, while the intervention elicited subjective improvements reported in the focus group, participants willing to participate were only the organisers, adding an extra layer of bias as their vested interest in the programme's success may have influenced their feedback. This potential bias in reporting could influence the validity of the perceived benefits of the intervention.

\section{Conclusion}

In this study, we designed and implemented a mindfulness-based course, specifically yoga, to explore the benefits of workplace well-being interventions in software engineer participants. Results from the quantitative analysis showed that the impact of yoga practice in this study was not statistically significant. It is essential to clarify that a lack of statistical significance does not imply that the intervention had no positive effects. Instead, it indicates that the observed changes could not be confidently attributed to the intervention based on the quantitative data. This may be due to a small sample size, participant response variability, or other uncontrolled variables. While statistical significance is a crucial marker for determining reliable effects, it is possible that the yoga practice had subtle or individual-level benefits that were not detected in the quantitative analysis.
Furthermore, the qualitative data from the focus group and the employee's feedback reported to the organisers was mainly positive. The yoga course is now an option for employees offered by the company, and they are still attending it. Employees might find other benefits that were not captured by the scales. Hence, they are still attending the course.

Future work will include making sure to control for confounding variables and to have longitudinal follow-up after intervention data collection.

\section*{Acknowledgment}
We thank the company and the managers who allowed us to run the intervention with them.

\bibliographystyle{ACM-Reference-Format}
\bibliography{bib.bib}

\vspace{12pt}
\color{red}

\end{document}